\documentclass[final,3p,times,twocolumn]{elsarticle}
\usepackage{amssymb}
\usepackage{color}
\journal{Physics Letters B}

\begin{document}
\begin{frontmatter}
\title{Measurement of very forward neutron energy spectra for 7 TeV proton--proton collisions at the Large Hadron Collider}
\author[label1,label2]{O.~Adriani}
\author[label1,label2]{E.~Berti}
\author[label1]{L.~Bonechi}
\author[label1,label2]{M.~Bongi}
\author[label1,label3]{G.~Castellini}
\author[label1,label2]{R.~D'Alessandro}
\author[label1,label2]{M.~Del~Prete}
\author[label4]{M.~Haguenauer}
\author[label5,label6]{Y.~Itow}
\author[label7]{K.~Kasahara}
\author[label5]{K.~Kawade\corref{cor1}}
\ead{kawade@stelab.nagoya-u.ac.jp}
\author[label5]{Y.~Makino}
\author[label5]{K.~Masuda}
\author[label5]{E.~Matsubayashi}
\author[label8]{H.~Menjo}
\author[label2]{G.~Mitsuka}
\author[label5]{Y.~Muraki}
\author[label5]{Y.~Okuno}
\author[label1]{P.~Papini}
\author[label9]{A-L.~Perrot}
\author[label1,label3]{S.~Ricciarini}
\author[label5,label6]{T.~Sako}
\author[label6]{N.~Sakurai}
\author[label5]{Y.~Sugiura}
\author[label7]{T.~Suzuki}
\author[label10]{T.~Tamura}
\author[label1,label2]{A.~Tiberio}
\author[label7]{S.~Torii}
\author[label11,label12]{A.~Tricomi}	
\author[label13]{W.C.~Turner}
\author[label5]{and Q.D.~Zhou}
\address[label1]{INFN Section of Florence, Italy}
\address[label2]{University of Florence, Italy}
\address[label3]{IFAC-CNR, Italy}
\address[label4]{Ecole-Polytechnique, France}
\address[label5]{Solar-Terrestrial Environment Laboratory, Nagoya University, Japan}
\address[label6]{Kobayashi-Maskawa Institute for the Origin of Particles and the Universe, Nagoya University, Japan}
\address[label7]{RISE, Waseda University, Japan}
\address[label8]{Graduate school of Science, Nagoya University, Japan}
\address[label9]{CERN, Switzerland}
\address[label10]{Kanagawa University, Japan}
\address[label11]{INFN Section of Catania, Italy}
\address[label12]{University of Catania, Italy}
\address[label13]{LBNL, Berkeley, USA}
\cortext[cor1]{corresponding author}

\begin{abstract}
The Large Hadron Collider forward (LHCf) experiment is designed to use the LHC to verify the hadronic-interaction models used in cosmic-ray physics.
Forward baryon production is one of the crucial points to understand the development of cosmic-ray showers.
We report the neutron-energy spectra for LHC $\sqrt{s}$ = 7 TeV proton--proton collisions with the pseudo-rapidity $\eta$ ranging from 8.81 to 8.99, from 8.99 to 9.22, and from 10.76 to infinity.
The measured energy spectra obtained from the two independent calorimeters of Arm1 and Arm2 show the same characteristic feature before unfolding the difference in the detector responses.
We unfolded the measured spectra by using the multidimensional unfolding method based on Bayesian theory, and the unfolded spectra were compared with current hadronic-interaction models.
The QGSJET II-03 model predicts a high neutron production rate at the highest pseudo-rapidity range similar to our results and the DPMJET 3.04 model describes our results well at the lower pseudo-rapidity ranges. 
However no model perfectly explains the experimental results in the whole pseudo-rapidity range.
The experimental data indicate the most abundant neutron production rate relative to the photon production, which does not agree with predictions of the models.
\end{abstract}


\end{frontmatter}

\section{Introduction}
The forward particle production process induced by collisions of high-energy particles is a poorly understood phenomenon in high-energy physics.
Though it is important to understand the development of cosmic-ray showers in the atmosphere, the validity of hadronic-interaction models have not been sufficiently verified at energies for ultra-high-energy cosmic rays (UHECRs, $>$ 10$^{18}$ eV) because of lack of experimental data in this energy range.
This lack of data results into a large uncertainty in the interpretation of the energy and chemical composition of UHECRs.
Forward baryons play a very important role in the development of cosmic-ray showers.
If forward baryons carry more collision energy, cosmic-ray showers develop much deeper in the atmosphere, and vice versa.
However, in the energy range of UHECRs, the predictions by current models differ significantly among themselves.

The excess of muons at ground level is reported as one of the problems in the cosmic-ray shower observations.
The number of muons observed by the surface detector array of the Pierre Auger Observatory (PAO) \cite{bib:PAO} is higher than the one expected based on the energy determined by the fluorescence detectors even if a heavy primary mass is assumed \cite{bib:muonexcess}.
It is suggested that the number of (anti) baryons generated in the forward region is strongly related to the number of muons observed by PAO at the ground \cite{bib:muon}.
Therefore, baryon production at the very forward region is quite important to understand cosmic-ray showers.

In this paper, we report the results of analyzing the data of the Large Hadron Collider forward (LHCf) experiment for forward neutron spectra.
Forward baryon spectra at the laboratory equivalent energy of 2.5$\times$10$^{16}$\,eV ($\sqrt{s}$=7\,TeV) will be a crucial input to improve 
the hadronic-interaction models used in the air shower analyses.

\section{LHCf experiment}

The LHCf experiment is designed to use the LHC to verify the hadronic-interaction models used in cosmic-ray experiments \cite{bib:LHCfTDR,bib:LHCfJINST}.
Two independent detectors named Arm1 and Arm2 were installed in the detector installation slots of the TANs located 140 m away from the interaction point 1 (IP1).
Because charged particles are swept away by the D1 bending magnets, LHCf can measure only neutral particles in the very forward region of the LHC (pseudo rapidity $|\eta| > 8.4$).
Both detectors have two different sampling calorimeters with 44 radiation lengths (1.6 hadron-interaction lengths) of tungsten plates and 16 layers of sampling scintillators \cite{bib:LHCfJINST}.
Four layers of the position sensors (SciFi in Arm1 and silicon micro-strip sensor in Arm2) can measure the hit position transverse to the beam direction.
The transverse dimensions of the calorimeters are 20 mm $\times$ 20 mm and 40 mm $\times$ 40 mm in Arm1, and 25 mm $\times$ 25 mm and 32 mm $\times$ 32 mm in Arm2.
The details of the detector performance during the 2009--2010 proton--proton collisions are reported in \cite{bib:LHCfIJMPA}.

The performance of the LHCf detectors for hadron measurements was studied by Monte Carlo (MC) simulations and confirmed by using 350 GeV proton beams at CERN-SPS \cite{bib:LHCfneutron}.
Depending on the incident-neutron energy, the energy resolution and position resolution are about 40\% and 0.1--1.3 mm, respectively.
The detection efficiency for neutrons was estimated to be 70\%--80\% for neutrons above 500 GeV.

In this paper we assume hadronic showers are produced by neutrons.
According to the EPOS 1.99 generator \cite{bib:EPOS}, about 10\% of other hadrons, i.e., $\Lambda$s and K$^{0}$s, are also included in the data and this fraction and the energy dependence are model dependent.

\section{Analysis}
\subsection{Data used in analysis}
The data used in this analysis were obtained on May 15, 2010 from proton--proton collisions at $\sqrt{s}$ = 7 TeV (LHC Fill \# 1104).
The typical luminosity corresponding to this fill derived from the counting rate of the LHCf front counters \cite{bib:LHCfFC} was (6.3--6.5) $\times 10^{28} {\rm cm}^{-2}{\rm s}^{-1}$.
The data set was the same as one used in the previously published photon analysis results, and additional details can be found in \cite{bib:LHCfphoton}.
The trigger for LHCf events was generated when signals from any three successive scintillation layers in any calorimeter exceeded a predefined threshold (typically 130 minimum ionizing particles (MIPs)).
Data acquisition (DAQ) was performed under 85.7\% (Arm1) and 67.0\% (Arm2) average live-times.

Taking the DAQ live-times into account, the integrated luminosities of the data set were 0.68 nb$^{-1}$ for Arm1 and 0.53 nb$^{-1}$ for Arm2, each with $\pm$6.1\% uncertainty.
The numbers of inelastic collisions were about 48M and 38M collisions for Arm1 and Arm2, respectively.

MC predictions were conducted with the generators DPMJET 3.04 \cite{bib:DPMJET3}, EPOS 1.99 \cite{bib:EPOS}, PYTHIA 8.145 \cite{bib:PYTHIA}, QGSJET II-03 \cite{bib:QGSJET2}, and SYBILL 2.1 \cite{bib:SYBILL} and compared with the experimental results.
In the MC simulations, the COSMOS (v7.49) and EPICS (v8.81) \cite{bib:EPICS} libraries that are used in air-shower and detector simulations were used to simulate the flight of particles from the IP1 to the detectors and the response of the detectors.
About 10M inelastic collisions were simulated for each model.

\subsection{Event reconstruction}\label{sec:reconst}
Initially, the offline-event selection was applied when energy depositions equivalent to more than 200 MIPs were recorded for three successive layers in addition to the experimental trigger.
The positions at which particles hit the detector were determined by using the position sensors.
Because reconstruction of the events is difficult at the edge of the calorimeters due to large fluctuations in the energy deposition, events within 2\,mm from the edge were discarded from the analysis.
The lateral shower leakage caused by the limited lateral size of the detectors deteriorates the energy resolution.
This position-dependent leakage effect was corrected by using the transverse-hit position measured by the position sensors.

Particle identification (PID) between neutrons and photons was based on the difference in the longitudinal shape of the shower development.
Two simple parameters called $L_{20\%}$ and $L_{90\%}$ were introduced to characterize the shower shape.
These parameters were defined as the depths containing 20\% and 90\%, respectively, of the total energy deposited within the layers.
Considering the correlation between $L_{20\%}$ and $L_{90\%}$ an optimized parameter L$_{{\rm 2D}}$ was defined as ${\rm L}_{{\rm 2D}} = L_{90\%} - 1/4 \times L_{20\%}$ to improve the selection efficiency and purity than previous analyses.
\begin{figure}[]
 \centering
 \includegraphics[width=0.45\textwidth]{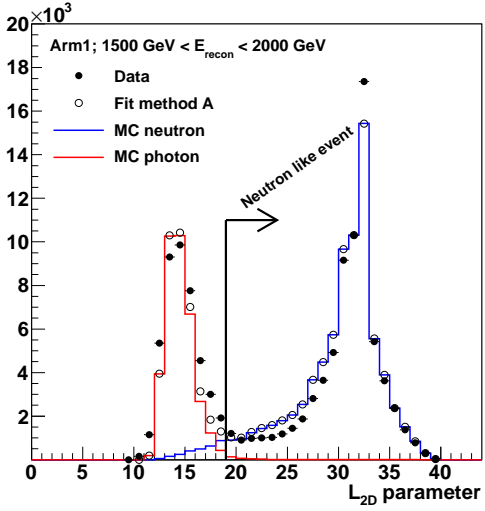}
 \caption{The L$_{{\rm 2D}}$ parameter distribution for the experimental data and the MC simulations from the template MC. The closed circles represent the Arm1 experimental results, whereas the red and blue histograms correspond to photon and neutron predictions. The open circles is the scaled results of the MC simulation obtained by method A.}
 \label{fig:L2D}
\end{figure}
Figure \ref{fig:L2D} shows the L$_{{\rm 2D}}$ distributions.
Two distinct peaks are identified in the observed L$_{{\rm 2D}}$ distribution indicated by the closed circles.
The histograms correspond to the MC prediction of pure photons (red) and pure neutrons (blue) and they are called `templates' hereafter.
The templates were produced by accumulating the MC simulation as a mixture of five models, DPMJET 3.04, EPOS 1.99, PYTHIA 8.145, QGSJET II-03, and SYBILL 2.1, with same statistics for each.

To obtain neutron spectra, only events with the parameter L$_{{\rm 2D}}$ exceeding a certain threshold were identified as neutron-like events.
The effects of the neutron selection efficiency and the photon contamination were corrected by using the efficiency $\epsilon$ and purity $P$ (i.e., multiplying by $P/\epsilon$ ) determined by MC simulations and the template fitting method \cite{bib:LHCfIJMPA,bib:LHCfphoton,bib:TFracFit}, respectively.
To estimate the photon contamination, the templates for photons and hadrons were independently scaled to reproduce the experimental results (called method A).
The open circles in Figure \ref{fig:L2D} represent the fitting result with method A.
Neutron selection criteria, as indicated in Figure \ref{fig:L2D}, were chosen to maximize $\epsilon \times P$.
To cope with energy dependence, events were separated into eight different energy ranges according to reconstructed energy.

After correcting the PID efficiency and purity, we obtain the production rate of neutrons as a function of obtained energy
that was determined from the total deposited energy in the calorimeter and from the identity of the particle.
Details of the event reconstruction for neutrons are summarized in \cite{bib:LHCfneutron}.
About 0.3 million neutron-like events passed the PID selection for each arm.

To combine the results of Arm1 and Arm2, we selected events that occurred within the common rapidity regions.
Events within 6 mm ($\eta > 10.76$) from the beam center were selected for the small towers.
The large towers of Arm1 and Arm2 were divided into two regions.
The inner region "A" was defined by a radius of 28--35 mm ($ 8.99 < \eta < 9.22$) from the beam center, whereas the outer region "B" was defined by a radius of 35--42 mm ($8.81 < \eta < 8.99$).
For the analysis, we used azimuthal-angle intervals  $d\phi$ of $360^{\circ}$ for the small towers and $20^{\circ}$ for the large towers.

\subsection{Systematic uncertainties}\label{sec:syserr}
\noindent{\bf Energy scale}\\
In order to determine energy scale and estimate its systematic uncertainty, we followed the previous analyses \cite{bib:LHCfphoton, bib:LHCfpizero}.
We observed invariant mass excesses of 8.1\% (Arm1) and 3.7\% (Arm2) compared with the $\pi^{0}$ mass reconstructed in the MC simulations.
These excesses were corrected in this analysis.
Based on this mass excess correction and known calibration uncertainty, totally $\pm$5.6\% (Arm1) and $\pm$4.4\% (Arm2) were assigned as systematic uncertainties with respect to the central value of the mass shift.
In addition, according to the differences between the SPS beam test and MC simulation in the reconstructed energy of 350 GeV proton showers \cite{bib:LHCfneutron}, +2.0\% (Arm1) and -3.8\% (Arm2) errors were quadratically added to the respective energy scale uncertainties.
\\ \\
{\bf PID}\\
Method A described in section \ref{sec:reconst} did not perfectly reproduce the experimental results.
To estimate systematic effects from these differences, we used a more artificial method (Method B) that allows longitudinal displacements and modifications in width of the distributions to fit the experimental results.
The systematic uncertainty from the PID process was estimated by comparing the results using the Method A and Method B.
The energy-dependent PID systematic uncertainty is mostly 1\% above 1.5 TeV and 12\% at most below this energy.
\begin{table*}[]
 \begin{center}
  \caption{Relative differences in eight bin energy spectra to evaluate systematic uncertainties from PID process. Small tower, Large tower A, and Large tower B correspond to the rapidity ranges from 10.76 to infinity, from 8.99 to 9.22, and from 8.82 to 8.99, respectively.}
  \begin{tabular}{c|rrrrrrrr}
   LHCf Energy   & \multicolumn{1}{l}{100} & \multicolumn{1}{l}{500} & \multicolumn{1}{l}{1000} & \multicolumn{1}{l}{1500} & \multicolumn{1}{l}{2000} & \multicolumn{1}{l}{2500} & \multicolumn{1}{l}{3000} &  \\ 
    {}[GeV]         & -500 & -1000 & -1500 & -2000 & -2500 & -3000 & -3500 & 3500$<$ \\ \hline 
   Small tower   & -8.0\% & -3.4\% & -1.8\% & -1.1\% & 0.0\% & 0.0\% & 0.0\% & -0.7\% \\
   Large tower A & -12.8\% & -9.7\% & -4.2\% & -1.7\% & -0.7\% & -0.6\% & 0.1\% & -0.8\% \\ 
   Large tower B & -17.7\% & -11.5\% & -3.6\% & -1.2\% & -0.7\% & -0.6\% & 0.4\% & -1.3\% \\
  \end{tabular}
  \label{tab:PIDerr}
 \end{center}
\end{table*}
The relative differences in the neutron production rate defined as
\begin{equation}
1-\frac{P_B/\epsilon_B}{P_A/\epsilon_A}
\end{equation}
are summarized in Table \ref{tab:PIDerr}.
Here, $\epsilon_A$ ($\epsilon_B$) and $P_A$ ($P_B$) are the efficiency and purity determined by Method A (Method B), respectively.
Final results are given using the Method A, while the differences shown in Table \ref{tab:PIDerr} are taken as a part of
the systematic uncertainties.
\\ \\
{\bf Multi-hit}\\
When two or more particles enter one of the LHCf calorimeters, these events are called `multihit' events.
Because discriminating between single and multihit events for neutron-like events was difficult because of large fluctuation of hadronic showers, rejection of multihit events causes a large systematic uncertainty.
Because the multihit event rate was predicted by MC simulations to be less than a few percent, all events used for this analysis were treated as single-hit events.
The systematic effects on the energy spectra from the multihit events were studied by using MC simulations.
We tested the difference of the spectra with current method and those with ideal multihit reconstruction by using the MC study.
The difference was less than 6\% above 500 GeV as summarized in Table \ref{tab:multi} and was taken into account as a
part of the systematic uncertainties.
\begin{table*}[]
 \begin{center}
  \caption{Relative differences of ten bin energy spectra to evaluate systematic error from multihit events. Small tower, Large tower A, and Large tower B correspond to the rapidity ranges from 10.76 to infinity, from 8.99 to 9.22, and from 8.82 to 8.99, respectively.}
  \begin{tabular}{c|cccccccccc}
   LHCf Energy & \multicolumn{1}{l}{100} & \multicolumn{1}{l}{500} & \multicolumn{1}{l}{1000} & \multicolumn{1}{l}{1500} & \multicolumn{1}{l}{2000} & \multicolumn{1}{l}{2500} & \multicolumn{1}{l}{3000} &  \multicolumn{1}{l}{3500} & \multicolumn{1}{l}{4000} &  \\ 
   {}[GeV] & -500& -1000& -1500& -2000& -2500& -3000& -3500 & -4000 & -5000 & 5000 $<$ \\ \hline
   Small tower  & 18.0\% & 5.6\% & 1.8 \% & 0.1 \% & 0.1 \% & 0.3 \% & -0.1\% & -2.5\% & -2.2\% &  -3.6\% \\
   Large tower A & 9.5 \% & 5.1\% & -0.4\% & 0.4 \% & 0.2 \% & -0.9\% & -2.5\% & -4.3\% & -4.0\% &  -2.4\% \\
   Large tower B & 8.8 \% & 2.3\% & 1.3 \% & -0.5\% & -0.3\% & -4.7\% & -2.4\% & -1.5\% & -6.5\% &  - \\
  \end{tabular}
  \label{tab:multi}
 \end{center}
\end{table*}
\\ \\
{\bf Position resolution}\\
Because the resolutions of transverse hit position for neutrons are 0.1--1.3\,mm depending on the neutron energy \cite{bib:LHCfneutron}, some events were reconstructed as hits at positions far from the true hit position.
Thus, particles hitting outside of the fiducial area may be contaminated and vice versa.
The effect of the position resolution was estimated by using MC simulations.

The difference between neutron-energy spectra selected by the reconstructed position and the true hit positions were calculated with EPOS 1.99, QGSJET II-03, SYBILL 2.1, DPMJET 3.04, and PYTHIA 8.145.
Because no significant dependence on model was found, the average of the predictions by the five different models was assigned to the systematic uncertainty.
The systematic error from these effects was less than 8.4\% at energies above 500 GeV and are summarized in Table \ref{tab:posreso}.
\begin{table*}[]
 \begin{center}
  \caption{Relative differences in eight bin energy spectra to evaluate systematic error due to position resolution. Small tower, Large tower A, and Large tower B correspond to the rapidity ranges from 10.76 to infinity, from 8.99 to 9.22, and from 8.82 to 8.99, respectively.}
  \begin{tabular}{c|cccccccc}
   LHCf Energy   & \multicolumn{1}{l}{100} & \multicolumn{1}{l}{500} & \multicolumn{1}{l}{1000} & \multicolumn{1}{l}{1500} & \multicolumn{1}{l}{2000} & \multicolumn{1}{l}{2500} & \multicolumn{1}{l}{3000} &  \\ 
   {}[GeV]                 & -500     & -1000   & -1500  & -2000   & -2500   & -3000   & -3500   & 3500 $<$ \\ \hline
   Small tower   & 23.2\%  & 8.4\%  & 4.5\% & 3.9\%  & 2.5\%  & 2.4\% & 1.9\% & 2.0\%  \\ 
   Large tower A & -54.2\% & -1.0\% & 0.9\% & -2.5\% & -0.6\% & -3.3\% & -3.6\% & -5.0\% \\
   Large tower B & -11.0\% & -0.0\% & 1.4\% & 0.6\%  & -1.1\% & -1.0\%   & -3.7\% & -5.7\% \\ 
  \end{tabular}
  \label{tab:posreso}
 \end{center}
\end{table*}
\\ \\
{\bf Other systematic errors}\\
As similar as the previous study \cite{bib:LHCfphoton}, the other systematic errors, such as the integrated luminosity ($\pm$6.1\%) and position of beam center (typically $\pm$3--10\%) were taken into account.
The systematic uncertainty from pile-up events (0.2\%) was negligibly small.



\section{Results}\label{sec:result}
\subsection{Measured energy spectra}
\begin{figure*}[]
 \centering
 \includegraphics[width=1.0\textwidth]{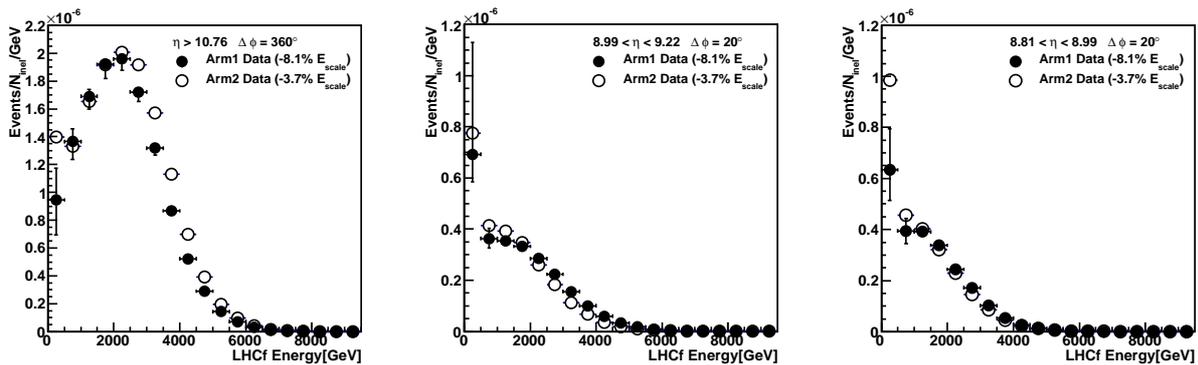}
 \caption{Energy spectra of neutron-like events measured by the Arm1 and Arm2 detectors. The left panel shows the results from the small towers, and the center and right panels show the results for the large towers. The horizontal axes represent the reconstructed energy. The vertical bars represent the statistical (they are negligibly small) and systematic uncertainties except the energy scale and the luminosity uncertainties.}
 \label{EnergySpectra}
\end{figure*}
Figure \ref{EnergySpectra} shows the energy spectra of forward neutrons measured by the LHCf Arm1 and LHCf Arm2 detectors.
The energy scale correction described above was applied in these spectra (-8.1\% for Arm1 and -3.7\% for Arm2).
The vertical axes were normalized to the number of inelastic collisions per GeV.
The vertical bars represent the statistical (they are very small) and systematic uncertainties except the energy scale and luminosity uncertainties \footnote{The uncertainty from the luminosity determination is not indicated in the figures throughout 
this paper.}.
The closed and open circles show the results of the Arm1 and Arm2 detectors, respectively.
A small difference in the detection efficiencies of the Arm1 and Arm2 detectors were not corrected here because it is treated in the
unfolding process discussed in Section \ref{sec-unfold}.
In spite of the difference in detector response, data from both Arms show the same characteristic feature of the spectra.

\begin{figure*}[]
 \centering
 \includegraphics[width=1.0\textwidth]{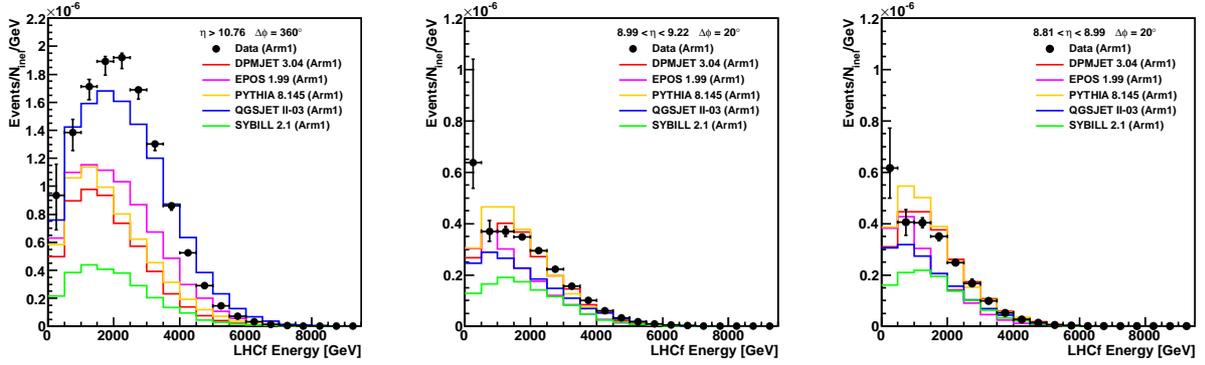}
 \caption{Measured Arm1 energy spectra of neutron-like events together with MC predictions. Left panel shows the results for the small tower, and the center and right panels show the results for the large tower. The vertical bars represent the statistical (they are very small) and systematic uncertainties except the energy scale and luminosity uncertainties. Colored lines indicate MC predictions by EPOS 1.99 (magenta), QGSJET II-03 (blue), SYBILL 2.1 (green), DPMJET 3.04 (red), and PYTHIA 8.145 (yellow).}
 \label{EnergySpectraA1}
\end{figure*}
Figures \ref{EnergySpectraA1} compares the energy spectra measured by the Arm1 detector with the MC predictions. 
Colored lines indicate MC predictions by EPOS 1.99 (magenta), QGSJET II-03 (blue), SYBILL 2.1 (green), DPMJET 3.04 (red), and PYTHIA 8.145 (yellow).
The model spectra were obtained from full detector simulations and taking account of the same reconstruction process as the experimental data.
None of the models perfectly matches the experimental data.
Experimental result indicates the hardest spectra than any other MC predictions.

\subsection{Spectra unfolding} \label{sec-unfold}
To estimate the true energy distribution, we used the multidimensional-spectra unfolding method \cite{bib:Unfold} with the variables energy and transeverse momentum ($p_T$).
To create training samples for the unfolding process, we used MC simulations with neutrons having a flat energy spectrum from 50 to 3500 GeV and an uniform injection to the detector plane.
The training samples were reconstructed by using the same method as the experimental data.
Performance of the unfolding method was checked by applying the unfolding process to the MC spectra and comparing the results with the true spectra.
\begin{figure}[h]
 \centering
 \includegraphics[width=0.45\textwidth]{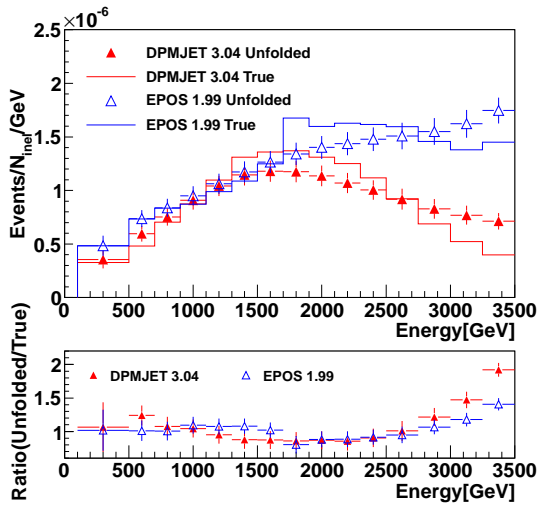}
 \caption{Comparison of unfolded spectra with true spectra for DPMJET 3.04 and EPOS 1.99 models at the small tower of Arm1. Bottom panel shows the ratio of the unfolded spectra to the true spectra.}
 \label{fig:Unfoldmodel}
\end{figure}
The upper panel in Figure \ref{fig:Unfoldmodel} shows the unfolded spectra for the DPMJET 3.04 and EPOS 1.99 models together with the true spectra at the small tower of Arm1.
Here "true spectra" means the true neutron energy distribution of the MC events after acceptance and trigger threshold were applied.
The bottom panel shows the ratio of the unfolded spectra to the true spectra.
The differences between the unfolded and true spectra were mostly within 20\% except at the highest energy bins (50-100\%).
These systematic differences were due to the choice of the flat energy distribution as a training sample.
We found that the difference did not strongly depend among five input models.
Thus we applied another correction; dividing the unfolded spectra by the average of differences.
The differences among the five models, typically $\pm$10\%,  were considered as a part of the systematic uncertainties.

\begin{figure*}[]
 \centering
 \includegraphics[width=1.0\textwidth]{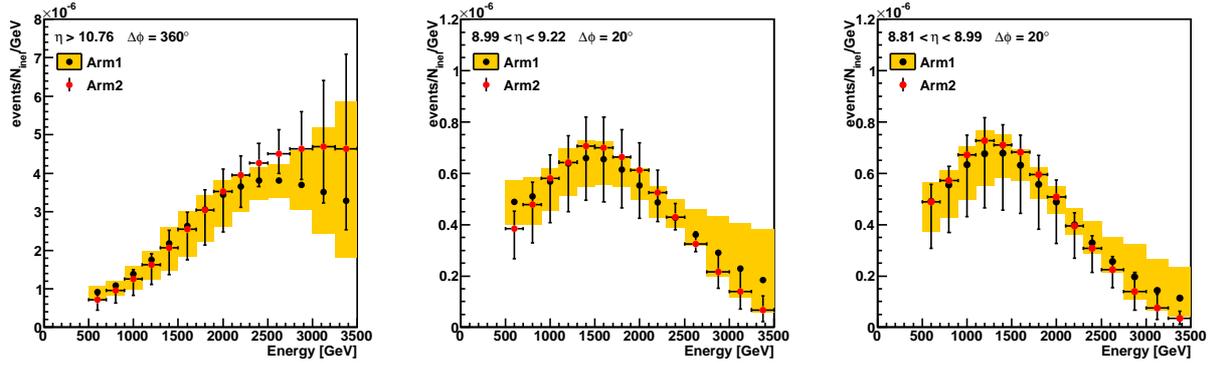}
 \caption{Unfolded energy spectra of the small towers ($\eta>10.76$) and the large towers ($8.99 < \eta<9.22$ and $8.81<\eta<8.99$). The hatched areas show the Arm1 systematic errors, and the bars represent the Arm2 systematic errors except the luminosity uncertainty..}
 \label{fig:Unfolddata}
\end{figure*}

Figure \ref{fig:Unfolddata} shows the unfolded experimental spectra measured by the Arm1 and Arm2 detectors for each rapidity range.
The hatched areas show the Arm1 systematic errors, and the bars represent the Arm2 systematic errors.
The detection efficiency of neutrons and the correction in the PID efficiency and purity were also considered.
The results below 500 GeV were not shown because of the large systematic errors.
The unfolded spectra from both Arms show good agreement within systematic errors. 

\begin{figure*}[]
 \centering
 \includegraphics[width=1.0\textwidth]{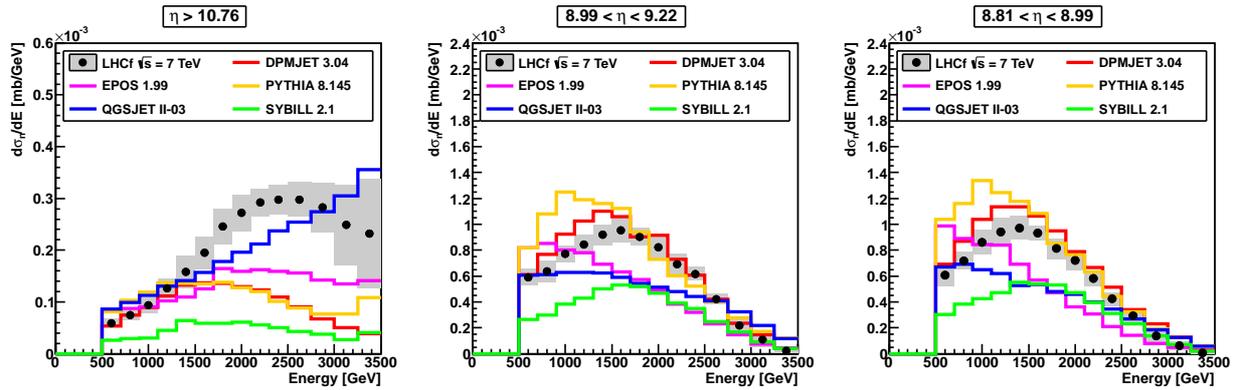}
 \caption{Comparison of the LHCf results with model predictions at small tower ($\eta>10.76$) and large towers ($8.99 < \eta<9.22$ and $8.81<\eta<8.99$). The black markers and gray hatched areas show the combined results of the LHCf Arm1 and Arm2 detectors and the systematic errors, respectively.}
 \label{fig:compmodel}
\end{figure*}
The experimental spectra were combined according to the previously used method \cite{bib:LHCfpizero}.
It was assumed that the systematic uncertainties caused by the energy scale, PID correction, beam center position, multihit events, and position resolution have bin-to-bin correlations while the other elements are independent between bins. 
The systematic uncertainties except the unfolding processes were thought to be fully uncorrelated between Arm1 and Arm2.
Because we treated the systematic uncertainties of the unfolding processes identical in Arm1 and Arm2, these errors were quadratically added after the combining process.

The differential neutron production cross sections $d\sigma_{n}/dE$ were calculated from the unfolded experimental spectra by using
\begin{equation}
d\sigma_{n}/dE = \frac{dN(\Delta\eta, \Delta E)}{\Delta E}\frac{1}{L} \times \frac{2\pi}{d\phi},
\end{equation}
where $dN(\Delta\eta, \Delta E)$ means the number of neutrons observed in the each rapidity range, $\Delta \eta$, and each energy bin, $\Delta E$. $L$ is the integrated luminosity corresponding to the data set.
The cross sections are summarized in Table \ref{tab:dr}.
Figure \ref{fig:compmodel} shows the combined Arm1 and Arm2 spectra together with the model predictions.
The experimental results indicate that the highest neutron production rate compared with the MC models occurs at the most forward rapidity.
The QGSJET II-03 model predicts a neutron production rate similar to the experimental results at the largest rapidity range. 
However, the DPMJET 3.04 model predicts neutron production rates better at the smaller rapidity ranges.

The neutron-to-photon ratios ($N_{n}/N_{\gamma}$) at three different rapidity regions are extracted after unfolding and summarized in Table \ref{tab:photohad}.
\begin{table}[]
 \begin{center}
  \caption{Hadron-to-photon ratio for experiment and MC models. The number of neutrons with energy above 100 GeV was divided by the number of photons with energy above 100 GeV. The rapidity intervals corresponding to the small tower, Large tower A, and Large tower B are $\eta > 10.76$, $9.22 > \eta > 8.99$, and $8.99 > \eta > 8.81$, respectively.}
  \begin{tabular}{c|ccc}
   $N_{n}/N_{\gamma}$        & Small & Large A & Large B \\ \hline
   Data         & 3.05$\pm$0.19 & 1.26$\pm$0.08 & 1.10 $\pm$0.07\\ \hline
   DPMJET 3.04  & 1.05 & 0.76 & 0.74\\
   EPOS 1.99    & 1.80 & 0.69 & 0.63\\
   PYTHIA 8.145 & 1.27 & 0.82 & 0.79\\
   QGSJET II-03 & 2.34 & 0.65 & 0.56\\
   SYBILL 2.1   & 0.88 & 0.57 & 0.53\\
  \end{tabular}
  \label{tab:photohad}
 \end{center}
\end{table}
Here $N_{n}$ and $N_{\gamma}$ are the number of neutrons and number of photons, respectively, with energies greater than 100\,GeV.
The numbers of photons were obtained from the previous analysis \cite{bib:LHCfphoton} and the same analysis for the pseudo-rapidity range 8.99-9.22 defined in this study.
The experimental data show the highest neutron ratio compared with the hadronic-interaction models for all rapidity regions.

\section{Summary and discussions}
An initial analysis of neutron spectra at the very forward region of the LHC is presented in this paper.
The data were acquired in May 2010 at the LHC from $\sqrt{s}$ = 7 TeV proton--proton collisions with integrated luminosity of 0.68 nb$^{-1}$ and 0.53 nb$^{-1}$ for the LHCf Arm1 and Arm2 detectors, respectively.

The neutron energy spectra were analyzed in the three different rapidity regions.
The results obtained from the two independent calorimeters of Arm1 and Arm2 are consistent each other.
The measured spectra were combined and unfolded by using a two-dimensional unfolding method based on Bayesian theory.

The experimental results, both in folded and unfolded were compared with the MC predictions of QGSJET II-03, EPOS 1.99, DPMJET 3.04, PYTHIA 8.145, and SYBILL 2.1; however, no model perfectly reproduces the experimental results in the whole pseudo-rapidity range.
Moreover, compared with the hadronic-interaction models, the experimental results show the most abundant neutron production relative to the photon production rate.

Because of the difference in the phase space, neutrons at lower rapidity carry more energy than those at higher rapidity. 
EPOS and QGSJET II 
\footnote{EPOS-LHC and QGSJET II-04 are the updated versions tuned with the initial LHC data.   
However the forward particle spectra do not have apparent difference from the model spectra shown in this paper.}, 
current standard models in the air shower analysis, underestimate the neutron production to about 30\% at the lower rapidity regions.
It is interesting to investigate how these differences affect to the development of air showers.

The differential cross sections for neutron production at very forward rapidity were measured at the Intersecting Storage Ring for proton--proton collisions at $\sqrt{s}$ = 30.6--62.7 GeV \cite{bib:ISR, bib:ISR2} and by the PHENIX experiment at the Relativistic Heavy Ion Collider for proton--proton collisions at $\sqrt{s}$ = 200 GeV \cite{bib:PHENIX}.
The results of previous experiments were consistent with Feynman x ($x_{F}$) scaling and do not depend on the collision energy.
To accurately extrapolate the hadronic-interaction models to the high-energy range (independently of whether or not the $x_{F}$ scaling in the neutron production is still relevant at energies in the TeV range) is also an important issue.
LHCf will extend the energy range to test the Feynman scaling with the proton--proton collision data obtained at $\sqrt{s}$= 0.9, 2.76, 7\,TeV and to be obtained at $\sqrt{s}$=13\,TeV in 2015. 




\section*{Acknowledgments}
We would like to express our gratitude to the CERN staff for their essential contribution to the operation of LHCf.
This study was supported by Grant-in-Aids for Scientific Research by MEXT of Japan, by the Grant-in-Aid for Nagoya University GCOE "QFPU" from MEXT, and by the Istituto Nazionale di Fisica Nucleare (INFN) in Italy.
Part of this work was performed using computer resources provided by the Institute for Cosmic-Ray Research at the University of Tokyo and by CERN.

\appendix
\section{Cross section table}
\begin{table*}[]
 \begin{center}
  \caption{Differential neutron production rate d$\sigma_{n}$/dE [mb/GeV] for each rapidity range.}
  \begin{tabular}{c|ccc}
   \multicolumn{4}{c}{Cross section [mb/GeV]}\\ \hline
   Energy [GeV] & Small tower ($\eta>10.76$)           & Large tower A ($8.99 < \eta<9.22$)    & Large tower B ($8.81<\eta<8.99$) \\ \hline
   500-700      & (5.91$\pm$0.81)$\times 10^{-5}$     & (5.91$\pm$0.66)$\times 10^{-4}$        & (6.09$\pm$0.84)$\times 10^{-4}$   \\
   700-900      & (7.48$\pm$0.95)$\times 10^{-5}$     & (6.38$\pm$0.82)$\times 10^{-4}$        & (7.15$\pm$0.67)$\times 10^{-4}$   \\
   900-1100     & (9.32$\pm$1.54)$\times 10^{-5}$    & (7.70$\pm$0.66)$\times 10^{-4}$        & (8.60$\pm$0.92)$\times 10^{-4}$   \\
   1100-1300    & (1.26$\pm$0.19)$\times 10^{-4}$   & (8.41$\pm$0.75)$\times 10^{-4}$        & (9.42$\pm$1.05)$\times 10^{-4}$   \\
   1300-1500    & (1.58$\pm$0.32)$\times 10^{-4}$   & (9.18$\pm$0.80)$\times 10^{-4}$        & (9.71$\pm$0.91)$\times 10^{-4}$   \\
   1500-1700    & (1.95$\pm$0.31)$\times 10^{-4}$   & (9.54$\pm$0.92)$\times 10^{-4}$        & (9.34$\pm$0.59)$\times 10^{-4}$   \\
   1700-1900    & (2.45$\pm$0.35)$\times 10^{-4}$   & (9.03$\pm$0.68)$\times 10^{-4}$        & (8.12$\pm$0.78)$\times 10^{-4}$   \\
   1900-2100    & (2.72$\pm$0.36)$\times 10^{-4}$   & (8.21$\pm$0.81)$\times 10^{-4}$        & (7.19$\pm$0.68)$\times 10^{-4}$   \\
   2100-2300    & (2.92$\pm$0.27)$\times 10^{-4}$   & (6.90$\pm$0.82)$\times 10^{-4}$        & (5.81$\pm$0.55)$\times 10^{-4}$   \\
   2300-2500    & (2.98$\pm$0.28)$\times 10^{-4}$   & (6.17$\pm$0.52)$\times 10^{-4}$        & (4.25$\pm$0.53)$\times 10^{-4}$   \\
   2500-2750    & (2.98$\pm$0.34)$\times 10^{-4}$   & (4.21$\pm$0.44)$\times 10^{-4}$        & (2.94$\pm$0.54)$\times 10^{-4}$   \\
   2750-3000    & (2.82$\pm$0.48)$\times 10^{-4}$   & (2.20$\pm$0.68)$\times 10^{-4}$        & (1.39$\pm$0.65)$\times 10^{-4}$   \\
   3000-3250    & (2.49$\pm$0.78)$\times 10^{-4}$   & (1.10$\pm$0.55)$\times 10^{-4}$        & (6.07$\pm$3.39)$\times 10^{-5}$   \\
   3250-3500    & (2.32$\pm$1.06)$\times 10^{-4}$   & (2.45$\pm$1.70)$\times 10^{-5}$        & (5.75$\pm$3.76)$\times 10^{-6}$   \\
  \end{tabular}
  \label{tab:dr}
 \end{center}
\end{table*}

\begin{thebibliography}{00}
\bibitem{bib:PAO} J.~Abraham, et al. Phys. Rev. Lett. {\bf 101} 061101 (2008).
\bibitem{bib:muonexcess} J.~Allen, et al. Proceedings of 32nd ICRC, Beijin (2011).
\bibitem{bib:muon} T.~Pierog, K.~Werner Phys. Rev. Lett. {\bf 101}, 171101 (2008).

\bibitem{bib:LHCfTDR} O.~Adriani, et al., LHCf-TDR, CERN-LHCC-2006-004.
\bibitem{bib:LHCfJINST} O.~Adriani, et al., JINST, {\bf 3}, S08006 (2008).
\bibitem{bib:LHCfIJMPA} O.~Adriani, et al., Int. J. Mod. Phys. A {\bf 28} 1330036 (2013).
\bibitem{bib:LHCfneutron} K.~Kawade, et al., JINST {\bf 9}, P03016 (2014).

\bibitem{bib:EPOS} K.~Werner, F.~M.~Liu and T.~Pierog, Phys. Rev. C, {\bf 74}, 044902 (2006).
\bibitem{bib:LHCfFC} K.~Taki et al., JINST {\bf 7}, T01003 (2012).
\bibitem{bib:LHCfphoton} O.~Adriani, et al., Phys. Lett. B, {\bf 703}, 128 (2011).
\bibitem{bib:DPMJET3} F.~W.~Bopp, J.~Ranft R.~Engel and S.~Roesler, Phys. Rev. C, {\bf 77}, 014904 (2008).
\bibitem{bib:PYTHIA} T.~Sj\"oostand, S.~Mrenna and P.~Skands, Comput. Phys. Comm., {\bf 178}, 852 (2008).
\bibitem{bib:QGSJET2} S.~Ostapchenko, Phys. Rev. D, {\bf 74}, 014026 (2006).
\bibitem{bib:SYBILL} E.-J.~Ahn, R.~Engel, T.~K.~Gaisser, P.~Lipari and T.~Stanev, Phys. Rev. D, {\bf 80}, 094003 (2009).
\bibitem{bib:EPICS} K.~Kasahara, EPICS web page, http://cosmos.n.kanagawa-u.ac.jp/
\bibitem{bib:TFracFit} R.~Barlow and C. Beeston, Comp. Phys. Comm. {\bf 77}, 219-228 (1993).
\bibitem{bib:LHCfpizero} O.~Adriani, et al., Phys. Rev. D, {\bf 86}, 092001 (2012).
\bibitem{bib:Unfold} G.~D'~Agostini, Nucl. Instrum. Meth. A {\bf 362}, 487 (1995).
\bibitem{bib:ISR} J.~Engler et al., Nucl. Phys. B {\bf 84}, 70 (1975).
\bibitem{bib:ISR2} W.~Flauger and F.~M\"{o}nnig, Nucl. Phys. B, {\bf 109}, 347-356 (1976).
\bibitem{bib:PHENIX} A.~Adare, et al., Phys. Rev. D, {\bf 88}, 032006 (2013).
\end{thebibliography}
\end{document}